\documentclass[12pt]{article}

\usepackage{amsmath}
\usepackage{amssymb}
\usepackage{cite}

\def\nn{\nonumber}

\numberwithin{equation}{section}

\title{Hamiltonian formulation for perfect fluid equations\\
 with the $\ell$--conformal Galilei symmetry}

\author{Timofei Snegirev${}^{a}$\thanks{timofei.v.snegirev@tusur.ru}
\\[0.5cm]
\it{\small ${}^a$Laboratory of Applied Mathematics and Theoretical Physics,}\\
\it{\small Tomsk State University of Control Systems and Radioelectronics,}\\
\it{\small Lenin ave. 40, 634050 Tomsk, Russia}}

\date{}

\begin{document}

\maketitle

\begin{abstract}
The Hamiltonian formulation for perfect fluid equations with the
$\ell$-conformal Galilei symmetry is proposed. For an arbitrary
half-integer value of the parameter $\ell$, the Hamilton and
non-canonical Poisson brackets are found, in terms of which the
original higher derivative equations of motion take the conventional
Hamiltonian form. The full set of conserved charges is found and
their algebra is established.
\end{abstract}

\thispagestyle{empty}
\newpage
\setcounter{page}{1}

\section{Introduction}

Conformal field theories in various dimensions continue to attract
considerable attention in connection with the development of the
AdS/CFT-correspondence. In particular, nonrelativistic conformal
field theory has recently been successfully applied within the
context of condensed matter physics \cite{NS2010}. An important
peculiarity of the nonrelativistic version of the
AdS/CFT-correspondence is that some of systems can be realized in
the laboratory \cite{OHGGT2002}.

Nonrelativistic conformal groups include dilatation transformation,
under which temporal and spatial coordinates scale differently:
$t'=\lambda t$, $x'_i=\lambda^\ell x_i$. The quantity
$z=\frac{1}{\ell}$ is known as the dynamic critical exponent and it
is an important object in condensed matter physics. As is known, the
Galilei algebra admits a finite-dimensional conformal extension for
an arbitrary (half)integer value of the parameter $\ell$, provided
constant accelerations are introduced \cite{Henkel,NOR}. The cases
$\ell=1/2$ and $\ell=1$ are referred to as the Schr\"{o}dinger
algebra and the conformal Galilei algebra, respectively. The latter
follows from the relativistic conformal algebra in the
nonrelativistic limit (see e.g. \cite{LSZ}).

Many physically interesting classical or quantum mechanics models
exhibit the $\ell=1/2$ conformal Galilei symmetry. A textbook
example is the Schr\"{o}dinger equation for a free massive particle
\cite{Nied1972,Hag1972}. Dynamical realizations with $\ell>1/2$ have
been the focus of extensive resent studies (see e.g.
\cite{LSZ1,FIL,DH1,GK,AGM,GM3,GM1,AGKM,AGGM,GM4,CG,M,KLS}). A
characteristic feature of such systems is the appearance of higher
derivative terms in the equations of motion, the order of which
correlates with the value of the parameter $\ell$.

Higher derivative theories are widely used in theoretical physics,
starting with radiation phenomena in classical electrodynamics
\cite{Gal02} and ending with ultraviolet regularization
\cite{Thir50} in quantum field theory. In addition, theories with
higher derivatives appear in string theory and M-theory
\cite{FradTh86}, as well as in field theories connected to
non-commutative geometry \cite{DN01}.

As was recently demonstrated in \cite{Gal22a,Gal22b}, a higher
derivative analog of Euler's perfect fluid equations enjoys the
$\ell$-conformal Galilei symmetry, provided a specific equation of
state is chosen\footnote{For a review of conformal symmetries of
nonrelativistic fluid mechanics see \cite{HH1,HH2,DH,HZ} and and
references therein.}. The goal of this work is to construct
Hamiltonian formulation, which generalizes the $\ell=1/2$ analysis
in \cite{Mor98,JNPP}. A consistent Hamiltonian formulation allows
one to strictly define the total energy of a system under
consideration. It is also needed for constructing a quantum theory.
The model presented in this work provides an example of a consistent
higher derivative Hamiltonian field theory.

The work is organized as follows. In Sect. \ref{S1}, we review the
structure of the $\ell$-conformal Galilei group and the associated
Lie algebra and its action in nonrelativistic space-time. In Sect.
\ref{S2}, we discuss the perfect fluid equations with the
$\ell$-conformal Galileo symmetry along the lines in \cite{Gal22a}.
The corresponding Hamiltonian formulation is constructed in Sect.
\ref{S3} for the case of a half-integer $\ell$ . In Sect. \ref{S4},
conserved charges associated with the $\ell$-conformal Galilei
symmetry are found and their algebra under the Poisson bracket is
verified. In concluding Sect. \ref{S5}, we summarize the results and
discuss possible further developments.

\section{The $\ell$-conformal Galilei group and its algebra}\label{S1}

Consider a nonrelativistic space-time parameterized by temporal
variable $t$ and Cartesian coordinates $x_i$, $i=1,...,d$. The
$\ell$-conformal Galilei group contains a subgroup of
$SL(2,R)$-transformations
\begin{eqnarray}\label{SL2}
t'=\frac{\alpha t+\beta}{\gamma t+\delta}, \quad
x'_i=\left(\frac{\partial t'}{\partial t}\right)^{\ell}x_i,
\end{eqnarray}
where $\alpha\delta-\beta\gamma=1$ and $\ell$ is an arbitrary
(half)integer parameter. These transformations describe
temporal translations, dilatations, and special conformal
transformations. The generators of the corresponding infinitesimal
transformations read
\begin{eqnarray}
H=\frac{\partial}{\partial t},\quad D=t\frac{\partial}{\partial
t}+\ell x_i\frac{\partial}{\partial x_i},\quad
K=t^2\frac{\partial}{\partial t}+2\ell tx_i\frac{\partial}{\partial
x_i},
\end{eqnarray}
which satisfy the commutation relations
\begin{eqnarray}\label{sl2}
{[H,D]}=H, \qquad {[H,K]}=2D,\qquad {[D,K]}=K.
\end{eqnarray}

The $\ell$-conformal Galilei group  also contains a chain of
transformations with vector parameters $a_i^{(n)}$ (no sum over $n$)
\begin{eqnarray}\label{Acc}
t'=t, \quad x'_i=x_i+a^{(n)}_it^n
\end{eqnarray}
where $n=0,1,...,2\ell$. For $n=0$ and $n=~1$ these transformations
correspond to spatial translations and Galilei boosts. For $n>1$ the
so-called constant accelerations show up. In infinitesimal form,
these transformations look similar, and their generators
\begin{eqnarray}
C^{(n)}_i=t^n\frac{\partial}{\partial x_i}
\end{eqnarray}
obey the non-vanishing commutators
\begin{eqnarray}\label{acc}
{[H,C^{(n)}_i]}=nC^{(n-1)}_i,\quad {[D,C^{(n)}_i]}=(n-l)C^{(n)}_i
\quad {[K,C^{(n)}_i]}=(n-2l)C^{(n+1)}_i.
\end{eqnarray}
The $\ell$-conformal Galilei group also involves  $SO(d)$ rotations which in what follows will be
disregarded.

Worth mentioning is that for a half-integer $\ell$ the algebra
admits a central extension \cite{GM11}
\begin{eqnarray}\label{CE}
[C^{(k)}_i,C^{(m)}_j]=(-1)^{k}k!m!\delta_{(k+m)(2\ell)}\delta_{ij} M,
\end{eqnarray}
where $M$ is the central charge usually attributed to a nonrelativistic mass.

\section{Equation of fluid dynamics with the $\ell$-conformal Galilei symmetry} \label{S2}

In hydrodynamics, a fluid is characterized by a density distribution
function $\rho(t,x)$ and the velocity vector field
$\upsilon_i(t,x)$, which both enter the continuity equation
\begin{eqnarray}\label{ConEq}
\frac{\partial\rho}{\partial t}+ \frac{\partial
(\rho\upsilon_i)}{\partial x_i}=0.
\end{eqnarray}
In the case of a perfect fluid,
the dynamics is described by the Euler equation
\begin{eqnarray}\label{EulEq}
{\cal D}\upsilon_i=-\frac{1}{\rho}\frac{\partial p}{\partial x_i},
\end{eqnarray}
where ${\cal D}=\frac{\partial}{\partial
t}+\upsilon_i\frac{\partial}{\partial x_i}$ is the material
derivative and $p(t,x)$ is the pressure. The latter is assumed to be
related to $\rho(t,x)$ via an equation of state. For a specific
equation of state, the continuity and Euler equations possess the
$\ell=\frac12$ conformal Galilei symmetry \cite{JNPP}. In order to
verify this, one needs to know transformation laws of $\rho(t,x)$
and $\upsilon_i(t,x)$ under the $\ell$-conformal Galilei group.

The transformation law of the density follows from the invariance of the mass
\begin{eqnarray}
\int_{V'}dx'\rho'(t',x')=\int_{V}dx\rho(t,x),
\end{eqnarray}
where a volume element $dx=dx_1dx_2...dx_d$ transforms as
$dx'=|\frac{\partial x'_i}{\partial x_j}|dx$. It gives
\begin{eqnarray}\label{SL2rho}
\rho(t,x)=\left(\frac{\partial t'}{\partial t}\right)^{\ell
d}\rho'(t',x')
\end{eqnarray}
for the $SL(2,R)$-transformations (\ref{SL2}) and
\begin{eqnarray}\label{Accrho}
\rho(t,x)=\rho'(t',x')
\end{eqnarray}
for the accelerations (\ref{Acc}).

In order to derive the transformation law of the velocity vector field
$\upsilon_i(t,x)$ it suffices to consider a particle orbit parameterized by $x_i(t)$ and take into account the relation
\begin{eqnarray}
\frac{dx_i(t)}{dt}=\upsilon_i(t,x(t)).
\end{eqnarray}
Differentiating (\ref{SL2}) and (\ref{Acc}) with respect to the
temporal variable, one finds
\begin{eqnarray}\label{SL2ups}
\upsilon_i(t,x)=x'_i\frac{\partial}{\partial t}\left(\frac{\partial
t'}{\partial t}\right)^{-\ell}+\left(\frac{\partial t'}{\partial
t}\right)^{1-\ell}\upsilon'_i(t',x')
\end{eqnarray}
for the $SL(2,R)$-transformations and
\begin{eqnarray}\label{Accups}
\upsilon_i(t,x)=\upsilon'_i(t',x')-na^{(n)}_it^{n-1}
\end{eqnarray}
for the accelerations.

One can show by direct calculations that the continuity and Euler
equations hold invariant under the transformations (\ref{SL2rho}),
(\ref{Accrho}) and (\ref{SL2ups}), (\ref{Accups}) for $\ell=1/2$,
provided the pressure $p(t,x)$ obeys the equation of state
\begin{eqnarray}
p=\nu \rho^{1+\frac{2}{ d}}
\end{eqnarray}
where $\nu$ is a real constant \cite{JNPP}.

In a recent work \cite{Gal22a}, the perfect fluid equations were
generalized to the case of an arbitrary value of $\ell$
\begin{eqnarray}\label{FlEql}
\frac{\partial\rho}{\partial t}+ \frac{\partial
(\rho\upsilon_i)}{\partial x_i}=0,\quad {\cal
D}^{2\ell}\upsilon_i=-\frac{1}{\rho}\frac{\partial p}{\partial
x_i},\quad p=\nu \rho^{1+\frac{1}{\ell d}}.
\end{eqnarray}
The energy-momentum tensor and conserved charges corresponding to
the $\ell$-conformal Galilei symmetry were constructed as well. In
particular, for a half-integer $\ell$
\begin{eqnarray}
\ell=n+1/2,\quad n=0,1,...
\end{eqnarray}
the conserved energy ${ H}=\int dxT^{00}$, which links to temporal translations, has the form
\begin{eqnarray}\label{Ham}
{ H}&=&\int dx \left[\frac12\rho\sum_{k=0}^{2n}(-1)^k{\cal
D}^{k}\upsilon_i{\cal D}^{2n-k}\upsilon_i+V(\rho)\right],
\end{eqnarray}
where $V(\rho)$ is a potential function, which via the Legendre
transformations gives the pressure
\begin{eqnarray}
p(\rho)=\rho V'(\rho)-V(\rho).
\end{eqnarray}
Given the equation of state $p=\nu \rho^{1+\frac{1}{\ell d}}$
\cite{Gal22a}, the potential reads
\begin{eqnarray}
V=\ell dp,
\end{eqnarray}
which also results in the identity
\begin{eqnarray}\label{Vcond}
\int dx(\ell x_i\frac{\partial p}{\partial x_i}+V)=0.
\end{eqnarray}
Our goal in the next section is to construct Hamiltonian formulation
for the $\ell$-conformal perfect fluid described by (\ref{FlEql}).

\section{Hamiltonian formulation} \label{S3}

As was demonstrated in \cite{JNPP}, for $\ell=\frac12$ the Hamilton
\begin{eqnarray}\label{HamPF}
H=\int dx\left[\frac12\rho\upsilon_i\upsilon_i+V(\rho)\right],\quad
V(\rho)=\frac12 d p
\end{eqnarray}
coincides with the conserved energy, provided the non-canonical
Poisson brackets for $\rho$ and $\upsilon_i$ are chosen
\begin{eqnarray}\label{PBPF}
\{\rho(x),\rho(y)\}&=&0,\nn
\\
\{\rho(x),\upsilon_i(y)\}&=&- \frac{\partial}{\partial
x_i}\delta(x-y),\nn
\\
\{\upsilon_i(x),\upsilon_j(y)\}&=&\frac{1}{\rho}\left(\frac{\partial\upsilon_j}{\partial
x_i} -\frac{\partial\upsilon_i}{\partial x_j}\right)\delta(x-y).
\end{eqnarray}
The Hamiltonian generates the continuity equation and the Euler
equation in the usual way
\begin{eqnarray}
\dot\rho=\{\rho,H\}=-\frac{\partial (\rho\upsilon_i)}{\partial
x_i},\qquad
\dot\upsilon_i=\{\upsilon_i,H\}=-\upsilon_j\frac{\partial\upsilon_i}{\partial
x_j}-\frac{1}{\rho}\frac{\partial p}{\partial x_i}
\end{eqnarray}
where the dot denotes the total time derivative.

In order to construct a similar Hamiltonian formulation for Eqs. (\ref{FlEql}),
we rewrite the equations
in the equivalent first order form
\begin{eqnarray}\label{HamFlEql}
\frac{\partial\rho}{\partial t}+ \frac{\partial
(\rho\upsilon^{(0)}_i)}{\partial x_i}=0,\quad {\cal
D}\upsilon^{(k)}_i=\upsilon^{(k+1)}_i ,\quad {\cal
D}\upsilon^{(2n)}_i=-\frac{1}{\rho}\frac{\partial p}{\partial x_i}
\end{eqnarray}
where $\upsilon^{(k)}_i$,
$k=0,1,...,2n$, are auxiliary fields with $\upsilon^{(0)}_i=\upsilon_i$.

Like in the $\ell=1/2$ case, it seems natural to choose the total energy (\ref{Ham}) to be the Hamilton
\begin{eqnarray}
H&=&\int dx
\left[\frac12\rho\sum_{k=0}^{2n}(-1)^k\upsilon^{(k)}_i\upsilon^{(2n-k)}_i+V(p)\right].
\end{eqnarray}
It remains to find Poisson brackets between $\rho$
and $\upsilon^{(k)}_i$ which represent (\ref{HamFlEql}) in the Hamiltonian form
\begin{eqnarray}
\dot\rho&=&\{\rho,H\}=-\frac{\partial
(\rho\upsilon^{(0)}_i)}{\partial x_i}\\
\dot\upsilon^{(k)}_i&=&\{\upsilon^{(k)}_i,H\}=-\upsilon^{(0)}_j\frac{\partial\upsilon^{(k)}_i}{\partial
x_j}+\upsilon^{(k+1)}_i
\\
\dot\upsilon^{(2n)}_i&=&\{\upsilon^{(2n)}_i,H\}=-\upsilon^{(0)}_j\frac{\partial\upsilon^{(2n)}_i}{\partial
x_j}-\frac{1}{\rho}\frac{\partial p}{\partial x_i}.
\end{eqnarray}
By direct inspection one finds the following brackets
\begin{eqnarray}\label{PBPFl}
&&\{\rho(x),\upsilon^{(k)}_i(y)\}=-\delta_{(k)(2n)}\frac{\partial}{\partial
x_i}\delta(x-y)
\\
\{\upsilon^{(k)}_i(x),\upsilon^{(m)}_j(y)\}&=&\frac{1}{\rho}\left(\delta_{(k)(2n)}\frac{\partial
\upsilon^{(m)}_j}{\partial x_i}- \delta_{(m)(2n)}\frac{\partial
\upsilon^{(k)}_i}{\partial
x_j}+(-1)^{k+1}\delta_{(k+m)(2n-1)}\delta_{ij}\right)\delta(x-y)\nn
\end{eqnarray}
where $\delta_{(k)(m)}$ is the Kronecker symbol. Note that there are
non-zero Poisson brackets between $\rho$ and $\upsilon^{(k)}_i$ for
$k=2n$, while the Poisson brackets between $\upsilon^{(k)}_i $ and
$\upsilon^{(m)}_i$ are non-zero for $k=2n$ and/or $m=2n$ or
$k+m=2n-1$. In order to verify the Jacobi identities, it suffices to
use the properties of delta-function
\begin{eqnarray}
&\delta(x-y)=\delta(y-x),\quad f(x)\delta(x-y)=f(y)\delta(x-y)\nn\\
& \dfrac{\partial}{\partial
x_i}\delta(x-y)=-\dfrac{\partial}{\partial y_i}\delta(x-y),
\end{eqnarray}
as well as their consequences
\begin{eqnarray}
\frac{\partial f(x)}{\partial x_i}\delta(x-y)=\frac{\partial
f(y)}{\partial y_i}\delta(x-y),\quad \frac{\partial^2}{\partial
x_i\partial y_j}\delta(x-y)&=&\frac{\partial^2}{\partial x_j\partial
y_i}\delta(x-y).
\end{eqnarray}
For $\ell=1/2$ the Hamiltonian formulation above correctly
reproduces that in \cite{JNPP}.

\section{The algebra of conserved charges}\label{S4}

Within the Hamiltonian framework, conserved charges obey a symmetry
algebra chosen under the Poisson bracket. In this section, we verify
this for the model at hand.

Switching to the first order formalism introduced above, one can rewrite conserved charges associated with the $SL(2,R)$-transformations \cite{Gal22a} in the following form
\begin{eqnarray}
H&=&\int dx
\left[\frac12\rho\sum_{k=0}^{2n}(-1)^k\upsilon^{(k)}_i\upsilon^{(2n-k)}_i+V(p)\right],
\\
D&=&tH-\frac12\int
dx\rho\sum_{k=0}^{2n}(-1)^{k}(k+1)\upsilon_i^{(k)}\upsilon_i^{(2n-k-1)},\nn
\\
K&=&t^2H-2tD-\frac12\int
dx\rho\sum_{k=0}^{2n}(-1)^k\left[(n+1)(2n+1)-k(k+1)\right]\upsilon^{(k-1)}_i\upsilon^{(2n-k-1)}_i,\nn
\end{eqnarray}
where $\upsilon^{(-1)}_i=x_i$. In order to verify that they are conserved, it suffices to use the relation
(\ref{Vcond}), the equations of motion
(\ref{HamFlEql}), and the following identities
\begin{eqnarray*}
\sum_{k=0}^{2n}(-1)^k\upsilon^{(k)}_i\upsilon^{(2n-k)}_i&=&{\cal
D}\sum_{k=0}^{2n}(-1)^k(k+1)\upsilon^{(k)}_i\upsilon^{(2n-k-1)}_i+\frac{1}{\rho}(1+2n)x_i\frac{\partial
p}{\partial x_i},
\\
\sum_{k=0}^{2n}(-1)^k(k+1)\upsilon^{(k)}_i\upsilon^{(2n-k-1)}_i&=&\frac12{\cal
D}\sum_{k=0}^{2n}(-1)^k[(n+1)(2n+1)-k(k+1)]\upsilon^{(k-1)}_i\upsilon^{(2n-k-1)}_i,
\\
\int dx\rho {\cal D}A&=&\frac{\partial}{\partial t}\int dx\rho A,
\end{eqnarray*}
where $A(t,x)$ is an arbitrary function. Computing the Poisson brackets, one reproduces the $sl(2,R)$ structure relations
\begin{eqnarray}
\{H,D\}=H, \qquad \{H,K\}=2D,\qquad \{D,K\}=K.
\end{eqnarray}

In order to establish the algebra of the acceleration generators
$C^{(k)}_i$, $k=0,1,...,2n+1$, for $\ell=n+\frac12$, it proves helpful to take into account the following identities
\begin{eqnarray}\label{Use}
\{\int dx \rho \upsilon^{(2n-k)}_i,H\}&=&\int dx \rho
\upsilon^{(2n-k+1)}_i,\quad k=1,...,2n+1
\end{eqnarray}
and $\{\int dx \rho \upsilon^{(2n)}_i,H\}=0$, which holds for $k=0$.
At this stage, let us introduce the quantities
\begin{eqnarray}
C_i^{(k)}&=&\sum_{s=0}^k\alpha(s)(-1)^st^{(k-s)}\int dx \rho
\upsilon^{(2n-s)}_i
\end{eqnarray}
with arbitrary coefficients $\alpha(s)$, and require that they are
conserved over time.
Using Eq. (\ref{Use}), one gets
\begin{eqnarray}
\dot{C}_i^{(k)}&=&\sum_{s=0}^{k-1}(-1)^s[\alpha(s)(k-s)-\alpha(s+1)]t^{(k-s-1)}\int
dx \rho \upsilon^{(2n-s)}_i,
\end{eqnarray}
which gives rise to the recurrence relation
\begin{eqnarray}
\alpha(s)(k-s)=\alpha(s+1) \quad \Rightarrow \alpha(s)=\frac{k!}{(k-s)!}\alpha(0),
\end{eqnarray}
$\alpha(0)$ being arbitrary normalization constant. Choosing $\alpha(0)=1$, one finally gets
\begin{eqnarray}
\{H,C^{(k)}_i\}=kC^{(k-1)}_i,\quad \{D,C^{(k)}_i\}=(k-l)C^{(k)}_i
\quad \{K,C^{(k)}_i\}=(k-2l)C^{(k+1)}_i,
\end{eqnarray}
as well as
\begin{eqnarray}
\{C_i^{(k)},C_j^{(m)}\}&=&(-1)^{k}k!m!\delta_{(k+m)(2n+1)}\delta_{ij}M,\quad
M=\int dx\rho,
\end{eqnarray}
where $M$ plays the role of the central charge \cite{GM11}.

\section{Conclusion}\label{S5}

To summarize, in this work the Hamiltonian formulation for the
perfect fluid equations with the $\ell$-conformal Galilei symmetry
was constructed. For an arbitrary half-integer value of the
parameter $\ell$, the Hamilton function was given and non-canonical
Poisson brackets were found, in terms of which the equations of
motion originally introduced in \cite{Gal22a} took the conventional
Hamiltonian form. Conserved charges associated with the
$\ell$-conformal Galilei group were constructed and their algebra
under the Poisson bracket was established. It was demonstrated that
within the Hamiltonian framework the algebra involved a central
charge $M$, which reflected the conservation of the mass. For
$\ell=1/2$, our results correctly reproduced the analysis in
\cite{JNPP}.

As a possible further development, it would be interesting to understand the
origin of the non-canonical Poisson bracket between the fundamental
fields. For $\ell=1/2$ and
three-dimensional space the canonical Poisson brackets can be
obtained by invoking to the Clebsch parametrization of the velocity vector
field \cite{JNPP}
\begin{eqnarray*}
\upsilon_i=\frac{\partial \theta}{\partial x_i}+\alpha
\frac{\partial \beta}{\partial x_i}
\end{eqnarray*}
with three scalar functions $\theta$, $\alpha$ and $\beta$. In this
parametrization the canonical variables will be $(\rho,\theta)$ and
$(\rho\alpha,\beta)$ with nontrivial Poisson brackets
\begin{eqnarray*}
\{\rho(x),\theta(y)\}=\delta(x-y),\quad
\{\alpha(x),\beta(y)\}=\frac{1}{\rho}\delta(x-y).
\end{eqnarray*}
Whether a similar parametrization is possible in any dimension
and for an arbitrary half-integer $\ell$ deserves a separate study.

\section*{Acknowledgements}
The author thanks A. Galajinsky for suggesting the problem, useful
discussions, and reading the manuscript. This work was supported by
the Russian Science Foundation, grant No 23-11-00002.


\begin{thebibliography}{10}


\bibitem{NS2010}
Y. Nishida, D.T. Son, {\it Unitary Fermi gas, epsilon expansion, and
nonrelativistic conformal field theories},  Lect. Notes Phys. {\bf
836} (2012) 233, arXiv:1004.3597.

\bibitem{OHGGT2002}
K. M. O'Hara, S. L. Hemmer, M. E. Gehm, S. R. Granade, J. E. Thomas,
{\it Observation of a strongly-interacting degenerate fermi gas of
atoms},  Science {\bf 298} (2002) 2179, arXiv:cond-mat/0212463.


\bibitem{Henkel}
M. Henkel, {\it Local scale invariance and strongly anisotropic
equilibrium critical systems}, Phys. Rev. Lett. {\bf 78} (1997)
1940, cond-mat/9610174.
\bibitem{NOR}
J. Negro, M.A. del Olmo, A. Rodriguez-Marco, {\it Nonrelativistic
conformal groups}, J. Math. Phys. {\bf 38} (1997) 3786.

\bibitem{LSZ}
J. Lukierski, P.C. Stichel, W.J. Zakrzewski, {\it Exotic Galilean conformal symmetry and its dynamical realisations}, Phys. Lett. A {\bf 357} (2006) 1, hep-th/0511259.


\bibitem{Nied1972}
U. Niederer, {\it The maximal kinematical invariance group of the
free Schr¨odinger equation}, Helv. Phys. Acta {\bf 45} (1972)
802-810.
\bibitem{Hag1972}
C. R. Hagen, {\it Scale and conformal transformations in
galilean-covariant field theory}, Phys. Rev. D {\bf 5} (1972)
377-388.




\bibitem{LSZ1}
J. Lukierski, P.C. Stichel, W.J. Zakrzewski, {\it
Acceleration-extended Galilean symmetries with central charges and
their dynamical realizations}, Phys. Lett. B {\bf 650} (2007) 203,
hep-th/0702179.

\bibitem{FIL}
S. Fedoruk, E. Ivanov, J. Lukierski, {\it Galilean conformal
mechanics from nonlinear realizations}, Phys. Rev. D {\bf 83} (2011)
085013, arXiv:1101.1658.

\bibitem{DH1}
C. Duval, P. Horvathy, {\it Conformal Galilei groups, Veronese
curves, and Newton--Hooke spacetimes}, J. Phys. A {\bf 44} (2011)
335203, arXiv:1104.1502.

\bibitem{GK}
J. Gomis, K. Kamimura, {\it Schrodinger equations for higher order
non-relativistic particles and N--Galilean conformal symmetry},
Phys. Rev. D {\bf 85} (2012) 045023, arXiv:1109.3773.

\bibitem{AGM}
K. Andrzejewski, J. Gonera, P. Maslanka, {\it Nonrelativistic
conformal groups and their dynamical realizations}, Phys. Rev. D
{\bf 86} (2012) 065009, arXiv:1204.5950.

\bibitem{GM3}
A. Galajinsky, I. Masterov, {\it Dynamical realization of
$\ell$--conformal Galilei algebra and oscillators}, Nucl. Phys. B
{\bf 866} (2013) 212, arXiv:1208.1403.

\bibitem{GM1}
A. Galajinsky, I. Masterov, {\it Dynamical realizations of
$\ell$--conformal Newton-Hooke group}, Phys. Lett. B {\bf 723}
(2013) 190, arXiv:1303.3419.

\bibitem{AGKM}
K. Andrzejewski, J. Gonera, P. Kosinski, P. Maslanka, {\it On
dynamical realizations of $\ell$--conformal Galilei groups}, Nucl.
Phys. B {\bf 876} (2013) 309, arXiv:1305.6805.

\bibitem{AGGM}
K. Andrzejewski, A. Galajinsky, J. Gonera, I. Masterov, {\it
Conformal Newton--Hooke symmetry of Pais-Uhlenbeck oscillator},
Nucl. Phys. B {\bf 885} (2014) 150, arXiv:1402.1297.



\bibitem{GM4}
A. Galajinsky, I. Masterov, {\it On dynamical realizations of
$\ell$--conformal Galilei and Newton--Hooke algebras}, Nucl. Phys. B
{\bf 896} (2015) 244, arXiv:1503.08633.

\bibitem{CG}
D. Chernyavsky, A. Galajinsky, {\it Ricci--flat spacetimes with
$\ell$--conformal Galilei symmetry}, Phys. Lett. B {\bf 754} (2016)
249, arXiv:1512.06226.


\bibitem{M}
I. Masterov, {\it Remark on higher-derivative mechanics with
$\ell$--conformal Galilei symmetry}, J. Math. Phys. {\bf 57} (2016)
092901, arXiv:1607.02693.

\bibitem{KLS}
S. Krivonos, O. Lechtenfeld, A. Sorin, {\it Minimal realization of
$\ell$--conformal Galilei algebra, Pais--Uhlenbeck oscillators and
their deformation}, JHEP {\bf 10} (2016) 078, arXiv:1607.03756.



\bibitem{Gal02}
D.V. Gal'tsov, {\it Radiation reaction in various dimensions}, Phys.
Rev. D 66 (2002) 025016, arXiv:hep-th/0112110.


\bibitem{Thir50}
W. Thirring, {\it Regularization as a consequence of higher order
equations}, Phys. Rev. 77 (1950) 570.

\bibitem{FradTh86}
E.S. Fradkin, A.A. Tseytlin, {\it Quantum string theory effective
action}, Nucl.Phys. B 261 (1985) 1-27.

\bibitem{DN01}
M.R. Douglas, N.A. Nekrasov, {\it Noncommutative field theory}, Rev.
Mod. Phys. 73 (2001) 977, arXiv:hep-th/0106048.

\bibitem{Gal22a}
A. Galajinsky, {\it Equations of fluid dynamics with the
$\ell$--conformal Galilei symmetry}, Nucl. Phys. B {\bf 984} (2022)
115965, arXiv:2205.12576.

\bibitem{Gal22b}
A. Galajinsky, {\it The group-theoretic approach to perfect fluid
equations with conformal symmetry}, Phys. Rev. D 107 (2023) 2,
026008, arXiv:2210.14544.


\bibitem{HH1}
M. Hassaine, P. A. Horvathy, {\it Field-dependent symmetries of a
non-relativistic fluid model}, Annals Phys. 282 (2000) 218-246,
arXiv:math-ph/9904022.

\bibitem{HH2}
M. Hassaine, P. A. Horvathy, {\it Symmetries of fluid dynamics with
polytropic exponent}, Phys.Lett. A279 (2001) 215-222,
arXiv:hep-th/0009092.

\bibitem{DH}
C. Duval, P. A. Horvathy, {\it Non-relativistic conformal symmetries
and Newton-Cartan structures}, J. Phys. A42 (2009) 465206,
arXiv:0904.0531.

\bibitem{HZ}
P.A. Horvathy, P.-M. Zhang, {\it Non--relativistic conformal
symmetries in fluid mechanics}, Eur. Phys. J. C {\bf 65} (2010) 607,
arXiv:0906.3594.







\bibitem{Mor98}
P. J. Morrison, {\it Hamiltonian description of the ideal fluid},
Rev. Mod. Phys. 70 (1998) 467.

\bibitem{JNPP}
R. Jackiw, V.P. Nair, S.Y. Pi, A.P. Polychronakos, {\it Perfect
fluid theory and its extensions}, J. Phys. A {\bf 37} (2004) R327,
arXiv:hep-ph/0407101.


\bibitem{GM11}
A. Galajinsky, I. Masterov, {\it Remarks on l-conformal extension of
the Newton-Hooke algebra}, Phys. Lett. B {\bf 702} (2011) 265-267,
arXiv:1104.5115.


\end{thebibliography}
\end{document}